\documentclass[aps,prl,twocolumn,showpacs,%
	amsmath,%
	reprint%
	,preprintnumbers,
	]{revtex4-1}

\usepackage{amsmath}
\usepackage{graphicx}
\usepackage{fancyhdr}

\begin{document}

\title{Quantum anholonomy with topology change}

\author{Taksu Cheon$^{*}$, Atushi Tanaka$^{\dagger}$, Ond{\v r}ej Turek$^{*}$}


\affiliation{
{}$^{*}$Laboratory of Physics, Kochi University of Technology, Tosa Yamada, Kochi 782-8502, Japan\\
{}$^{\dagger}$Department of Physics, Tokyo Metropolitan University, Hachioji, 192-0397 Tokyo, Japan}

\date{\today}

\begin{abstract}
We study a family of closed quantum graphs described by one singular vertex of order $n=4$.  By suitable choice of the parameters specifying the singular vertex, we can construct a closed sequence of paths in the parameter space that physically corresponds to the smooth interpolation of different topologies - a ring, separate two lines, separate two rings, 
two rings with a contact point.
We find that the spectrum of a quantum particle on this family of graphs shows the quantum anholonomy.
\end{abstract}

\maketitle
\thispagestyle{fancy}

\section{Introduction}

The idea of quantum mechanics on graphs has been introduced in the last century. During the last 30 years, it has been developed into a powerful tool for the study of particles on tiny graph-like structures \cite{EKST08}. Since a graph is a one-dimensional variety, the analysis of a particle on a graph consists in solving a system of \emph{ordinary} differential equations, which makes the graph models simple from the mathematical point of view. Therefore, quantum mechanics on graphs can serve also as a useful laboratory for the study of various complex quantum phenomena in a simple setting.

Here we examine the influence of the topology of quantum graphs on spectral properties, taking inspiration from the works \cite{BB95} and \cite{SW12}. We may regard that the topology change mimics a macroscopic or violent variation of systems, e.g., the rupture and connection of quantum wires. We will show that a parametric evolution of an eigenenergy may form an open trajectory along a cycle involving the topology changes of graphs. If this is the case, the trajectory of the corresponding eigenspace also forms an open cycle. Such a discrepancy of eigenobjects is called quantum anholonomy \cite{BE84, WZ84, CT09}.

\section{The Model}

\begin{figure}[h]
\center\includegraphics[width=4.0cm]{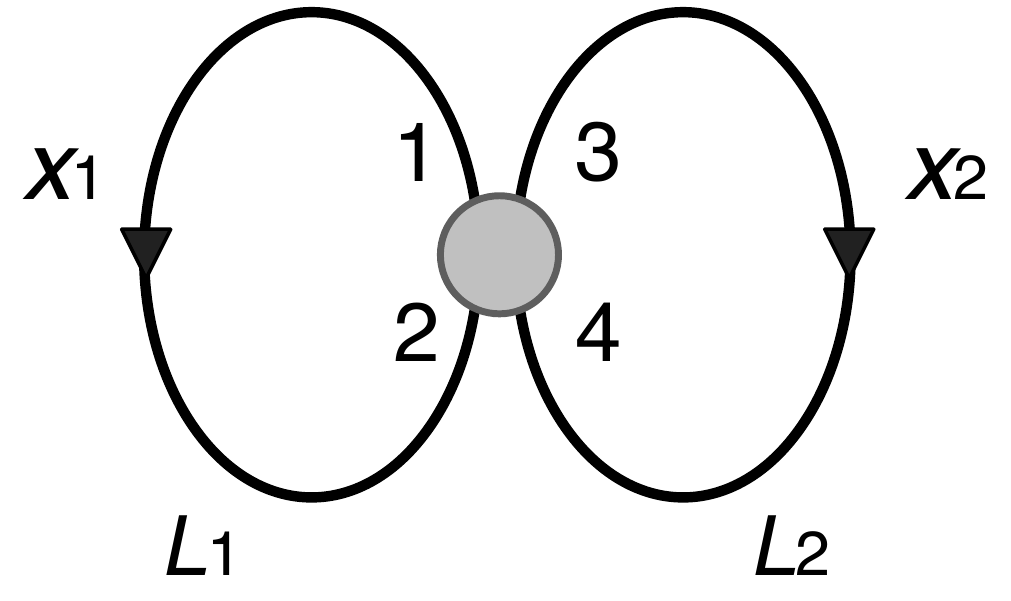}
\caption
{\label{fig1}
A closed quantum graph with an $n=4$ singular vertex. 
}
\end{figure}

Let us consider a singular node with four outgoing one-dimensional lines of finite lengths.   Let the first two lines be sticked together at their outer endpoints to form a ring of length $L_1$, and likewise, the second two lines be sticked together at their outer endpoints to form a ring of length $L_2$, giving the whole object the shape of the character ``$\infty$'' (Fig.~\ref{fig1}).  We assume there is no potential on the lines outside the central vertex. Therefore, a quantum particle confined to this object moves freely on the lines, and the only non-trivial part of the system is the connection condition of the wave function at the node. 

We assign the coordinates $x_1 \in [0,L_1]$ and $x_2 \in [0,L_2]$  to the left and the right ring. The values $x_1=0$ and $x_2=0$ correspond to the inner endpoints 1 and 3, respectively, whereas the values $L_1$ and $L_2$ correspond to the inner endpoints 2 and 4, see Fig.~\ref{fig1}.
The behavior of a particle in the central vertex of degree $4$ is determined by a boundary condition in the vertex. The general form of a boundary condition in a vertex of degree $n$ follows from the study of self-adjoint extensions of the Laplacian operator on a graph. The boundary condition consists of $n$ equations connecting the boundary values and derivatives that can be written in the form \cite{KS99}
\begin{eqnarray}
\label{ee1}
A \Psi + B \Psi' = 0 , 
\end{eqnarray} 
where $\Psi$, $\Psi'$ are vectors formed from the values of the wave function in the vertex and from its derivatives (taken in the outgoing sense), respectively, and $A$ and $B$ are $n\times n$ complex matrices with the property
\begin{eqnarray}
\label{ee2}
\quad AB^\dagger = B A^\dagger \qquad\text{and}\qquad \mathrm{rank}(A|B)=n
\end{eqnarray}
with $(A|B)$ denoting the $n\times2n$ matrix composed from the columns of the matrices $A$ and $B$.
For the graph depicted in Figure~\ref{fig1}, we have
\begin{eqnarray}
\Psi =  \begin{pmatrix} \psi_1(0) \\  \psi_1(L_1) \\ \psi_2(0)  \\ \psi_2(L_2) \end{pmatrix},
\quad
\Psi' =  \begin{pmatrix} \psi'_1(0) \\  -\psi'_1(L_1) \\ \psi'_2(0)  \\ -\psi'_2(L_2) \end{pmatrix} .
\end{eqnarray} 
In our model, we focus on a specific subset of all possible connection conditions (\ref{ee1}), namely
\begin{multline}\label{bc}
 \begin{pmatrix} 1 & t \, b(\theta)  & t \, a(\theta) & s \, d(\theta)\\
 0  & 1-c(\theta) & 0 &  t \, a(\theta) \\
 0  & 0 & c(\theta) & t \, b'(\theta) \\
 0  & 0& 0 & 0  \end{pmatrix} \Psi' \\ =
 \begin{pmatrix}  0& 0 & 0 & 0\\
  -t\, b(\theta) & c(\theta) & 0 & 0  \\
   -t\, a(\theta) & 0&1-c(\theta) & 0\\
    -s\, d(\theta)& -t\, a(\theta) & -t\, b'(\theta) & 1 \end{pmatrix} \Psi
\end{multline} 
where $t$ and $s$ are positive constants, and $a(\theta)$, $b(\theta)$, $b'(\theta)$, $c(\theta)$ and $d(\theta)$ are real periodic functions of $\theta$ with period $2\pi$ that vary within the range between $0$ and $1$.  In addition,
it can be easily shown that the functions must satisfy
\begin{eqnarray}
a(\theta) \ne 0 \ \  \Rightarrow \ \  c(\theta)=0, \\
b(\theta), b'(\theta) \ne 0 \ \  \Rightarrow \ \  c(\theta)=1
\end{eqnarray} 
for all $\theta\in[0,2\pi]$ in order that the matrices $A$, $B$ involved in the connection condition~\eqref{bc} obey (\ref{ee2}).   

\subsection{Long Cycle}
\begin{figure}[h]
\center\includegraphics[width=5.0cm]{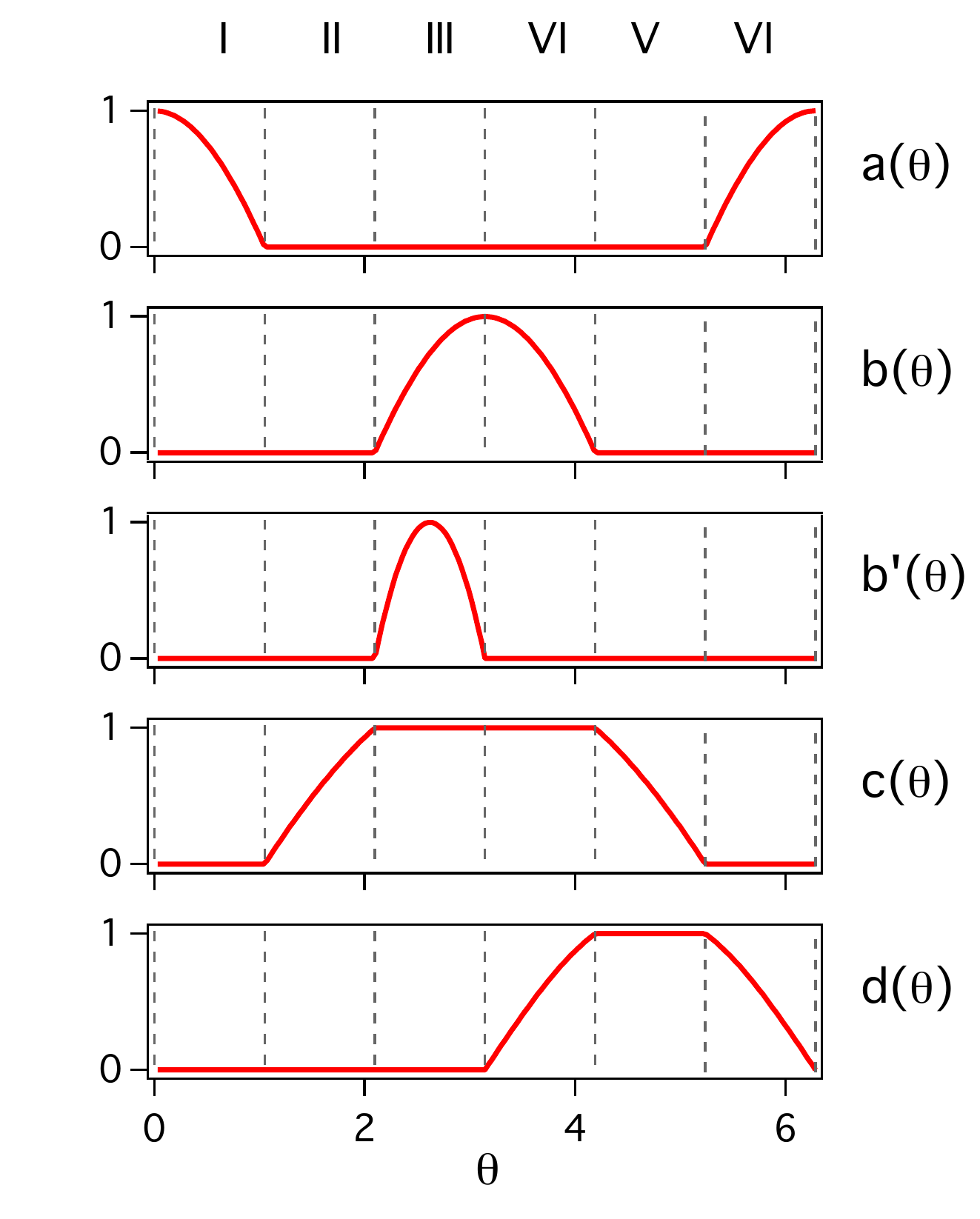}
\caption
{\label{fig2}
Graphs of functions defined by~\eqref{locyc} for $\theta\in[0,2\pi]$.
}
\end{figure}

At first we consider the boundary conditions~\eqref{bc} with functions $a(\theta)$, $b(\theta)$, $b'(\theta)$, $c(\theta)$ and $d(\theta)$ chosen as
\begin{align}
\label{locyc}
a(\theta) &= \cos\theta -\frac{1}{2} + \left| {\cos\theta -\frac{1}{2}} \right| ,
\nonumber \\
b(\theta) &= -\cos\theta -\frac{1}{2} + \left| {\cos\theta +\frac{1}{2}} \right| ,
\nonumber \\
b'(\theta) &= \frac{1}{4-2\sqrt{3}} \left\{ \sin\theta-\sqrt{3}\cos\theta -\sqrt{3} \right. \\
           & \qquad\qquad\qquad \left. - \left| {\sin\theta-\sqrt{3}\cos\theta -\sqrt{3}} \right| \right\} ,
\nonumber \\
c(\theta) &= \frac{1}{\sqrt{3}-1} 
 \left\{ 
   \frac{\sqrt{3}}{2} -\frac{1}{2} 
   + \left| { | \sin\frac{\theta}{2} | -\frac{1}{2} } \right| \right. \\
   & \qquad\qquad\qquad \left. - \left| { | \sin\frac{\theta}{2} | -\frac{\sqrt{3}}{2} } \right| 
 \right\} ,
\nonumber \\
d(\theta) &= \frac{1}{2\sqrt{3}} \left\{ |\sin\theta|-\sin\theta +\sqrt{3} \right. \\
          & \qquad\qquad\qquad \left. - \left| {|\sin\theta|-\sin\theta -\sqrt{3}} \right| \right\} .
\end{align} 
The graphs of these functions for $\theta\in[0,2\pi]$ are plotted in Fig.~\ref{fig2}.
The interval $[0,2\pi]$ can be divided into six regions, in each of them the vector $(a(\theta),b(\theta),b'(\theta),c(\theta),d(\theta))$ has a different structure of zero and nonzero entries. As we show below, these regions correspond to mutually different topologies of the system. It extends the basic idea of parametric systems with a varying topology introduced in works of Balachandran {\it et. al.} \cite{BB95} and also in  \cite{SW12}. Our use of a single $n=4$ singular vertex makes the treatment more systematic and unified.  

\begin{figure}[h]
\center{
\includegraphics[width=6.0cm]{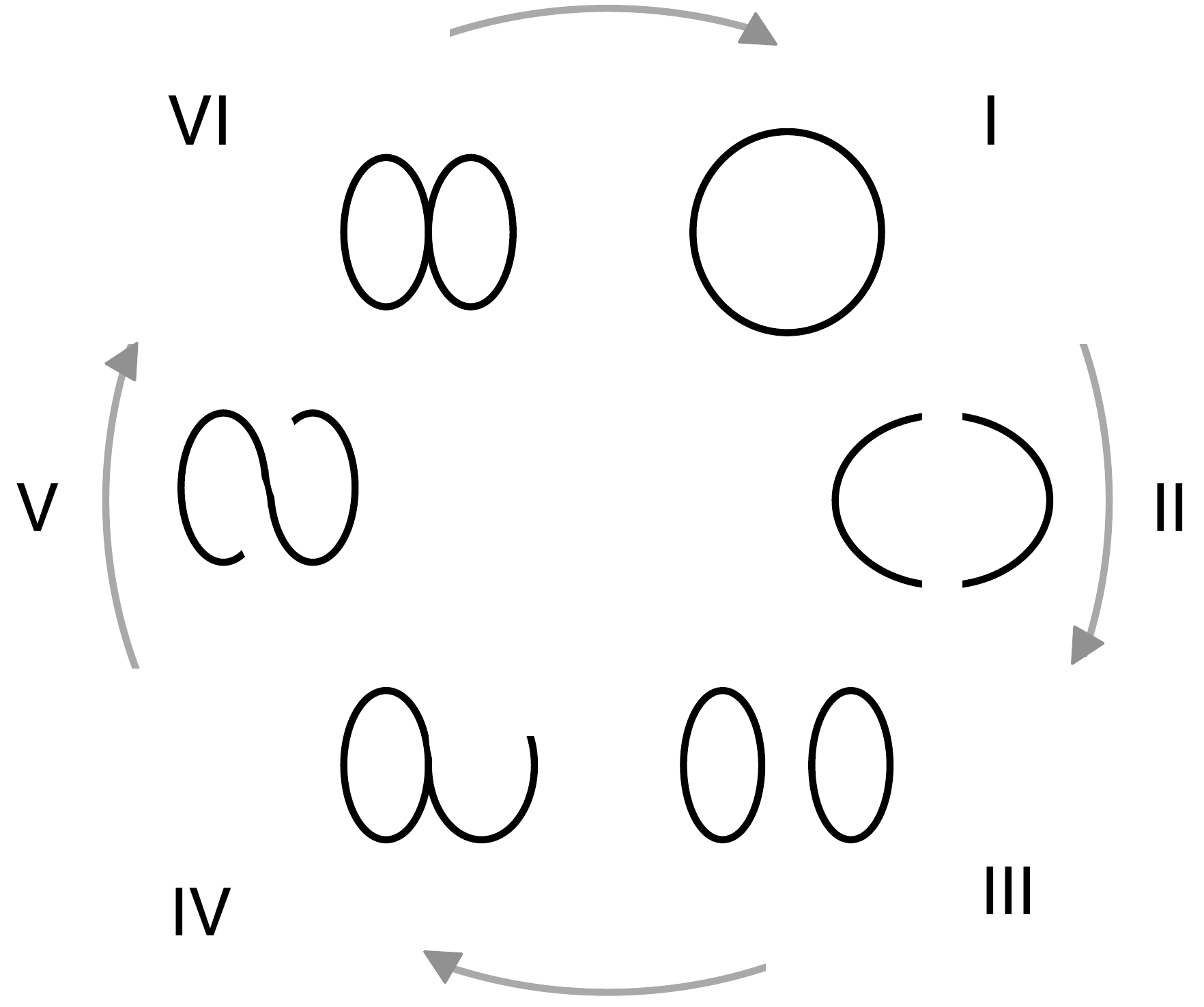}}
\caption
{\label{fig3}
Cycle of topological changes corresponding to functions~\eqref{locyc}.
}
\end{figure}

\paragraph*{Region I}
For $\theta\in[0,\frac{\pi}{3}]$, we have
\begin{multline}\label{I}
 \begin{pmatrix} 1 &  & t\cdot a(\theta) & \\  & 1 &  &  t\cdot a(\theta) \\  &  &  &  \\  & &  &   \end{pmatrix} \Psi' \\ =
 \begin{pmatrix}  &  &  & \\  &  &  &   \\ -t\cdot a(\theta) & &1 & \\ & -t\cdot a(\theta) & & 1 \end{pmatrix} \Psi ,
\end{multline} 
where $ta(\theta)$ takes the value $t$ at $\theta=0$ and it continuously decreases down to $0$ at $\theta= \frac{\pi}{3}$.
With regard to the structure of~\eqref{I}, the endpoint 1 is joined to the endpoint 2 and the endpoint 3 is joined to the edpoint 4. On the other hand, the endpoints 1 and 2 are disjoint, so are 3 and 4. Therefore, the system is topologically equivalent to a single ring, see Fig.~\ref{fig3}.

\paragraph*{Region II}
For $\theta\in[\frac{\pi}{3},\frac{2\pi}{3}]$, we have
\begin{multline}
 \begin{pmatrix} 1 &  &  & \\  & 1-c(\theta) &  &   \\  &  & c(\theta) &  \\  & &  &   \end{pmatrix} \Psi' \\ =
 \begin{pmatrix}  &  &  & \\  & c(\theta) &  &   \\  & & 1-c(\theta) & \\ &  & & 1 \end{pmatrix} \Psi ,
\end{multline} 
where we start with the value $c(\frac{\pi}{3})=0$  and continuously increase it up to $c(\frac{2\pi}{3})=1$.
Since all four inner endpoints are disjoint, the system is topologically equivalent to two separate lines.

\paragraph*{Region III}
For $\theta\in[\frac{2\pi}{3},\pi]$, we have
\begin{multline}\label{III}
 \begin{pmatrix} 1 & t\cdot b(\theta) &  & \\  &  &  &   \\  &  & 1 & t\cdot b'(\theta) \\  & &  &   \end{pmatrix} \Psi' \\ =
 \begin{pmatrix}  &  &  & \\  -t\cdot b(\theta) & 1 &  &   \\  & &  & \\ &  & -t\cdot b'(\theta) & 1 \end{pmatrix} \Psi ,
\end{multline} 
where $tb(\theta)$ continuously increases from $tb(\frac{2\pi}{3})=0$ up to $tb(\pi)=t$, while $tb'(\theta)$ takes the value $tb'(\theta)=0$ at $\theta=\frac{2\pi}{3}$, continuously increases up to $t$ at $\theta= \frac{5\pi}{6}$ and comes down back to $0$ at $\theta= {\pi}$.
The connection condition~\eqref{III} implies disjoints endpoints 1 and 3, as well as 2 and 4, making the system topologically equivalent to two separate rings.

\paragraph*{Region IV}
For $\theta\in[\pi,\frac{4\pi}{3}]$, we have
\begin{multline}
 \begin{pmatrix} 1 & t\cdot b(\theta) &  & s\cdot d(\theta)\\  &  &  &   \\  &  & 1 & \\  & &  &   \end{pmatrix} \Psi' \\ =
 \begin{pmatrix}  &  &  & \\  -t\cdot b(\theta)& 1 &  &   \\  & &  & \\ -s\cdot d(\theta) &  &  & 1 \end{pmatrix} \Psi ,
\end{multline} 
where $tb(\theta)$ continuously decreases from $tb(\frac{\pi}{2})=t$ to $tb(\frac{4\pi}{3})=0$, while
$sd(\theta)$ continuously increases from $sd(\frac{\pi}{2})=0$ up to $sd(\frac{4\pi}{3})=s$.
The system is topologically equivalent to a ring and a line connected at a single point since three of four endpoints, namely 1, 2 and 4, are connected in the node. Note that at the juncture of regions $III$ and $IV$, the system is topologically equivalent to a separated line and a ring.

\paragraph*{Region V}
For $\theta\in[\frac{4\pi}{3},\frac{5\pi}{3}]$, we have
\begin{multline}
 \begin{pmatrix} 1 &  &  & s\\  & 1-c(\theta) &  &   \\  &  & c(\theta) &  \\  & &  &   \end{pmatrix} \Psi' \\ =
 \begin{pmatrix}  &  &  & \\  & c(\theta) &  &   \\  & &1-c(\theta) & \\ -s &  & & 1 \end{pmatrix} \Psi ,
\end{multline} 
where $c$ continuously decreases from $c(\frac{4\pi}{3})=1$ down to $c(\frac{5\pi}{3})=0$.
The system is topologically equivalent to a single line, since only endpoints 1 and 4 are connected, whereas 2 and 3 are disjoint.

\paragraph*{Region VI}
For $\theta\in[\frac{5\pi}{3},2\pi]$, we have 
\begin{multline}
 \begin{pmatrix} 1 &  & t\cdot a(\theta) & s\cdot d(\theta)\\  & 1 &  &  t\cdot a(\theta) \\  &  &  &  \\  & &  &   \end{pmatrix} \Psi' \\ =
 \begin{pmatrix}  &  &  & \\  &  &  &   \\ -t\cdot a(\theta)& &1 & \\-s\cdot d(\theta) & -t\cdot a(\theta) & & 1 \end{pmatrix} \Psi ,
\end{multline} 
where $ta(\theta)$ continuously increases from $ta(\frac{5\pi}{3})$ up to $ta(2\pi)=t$, while
$sd(\theta)$ continuously decreases from $sd(\frac{5\pi}{3})=s$ down to $sd(2\pi)=0$.
In this region, the system is topologically equivalent to two rings connected at a single point since all four endpoints are connected together at the central node.

%
%


Therefore, with the parameter change $\theta = 0 \to 2\pi$, seven distinct topologies are traversed in a continuous manner. The situation is illustrated in Fig.~\ref{fig3}.
The system is never identical for any two different values of $\theta$ in $[0,2\pi)$, because the the functions $a(\theta)$ to $d(\theta)$ form district combinations of patterns in all six regions.

\subsection{Short Cycle}
\begin{figure}[h]
\center{
\includegraphics[width=5.0cm]{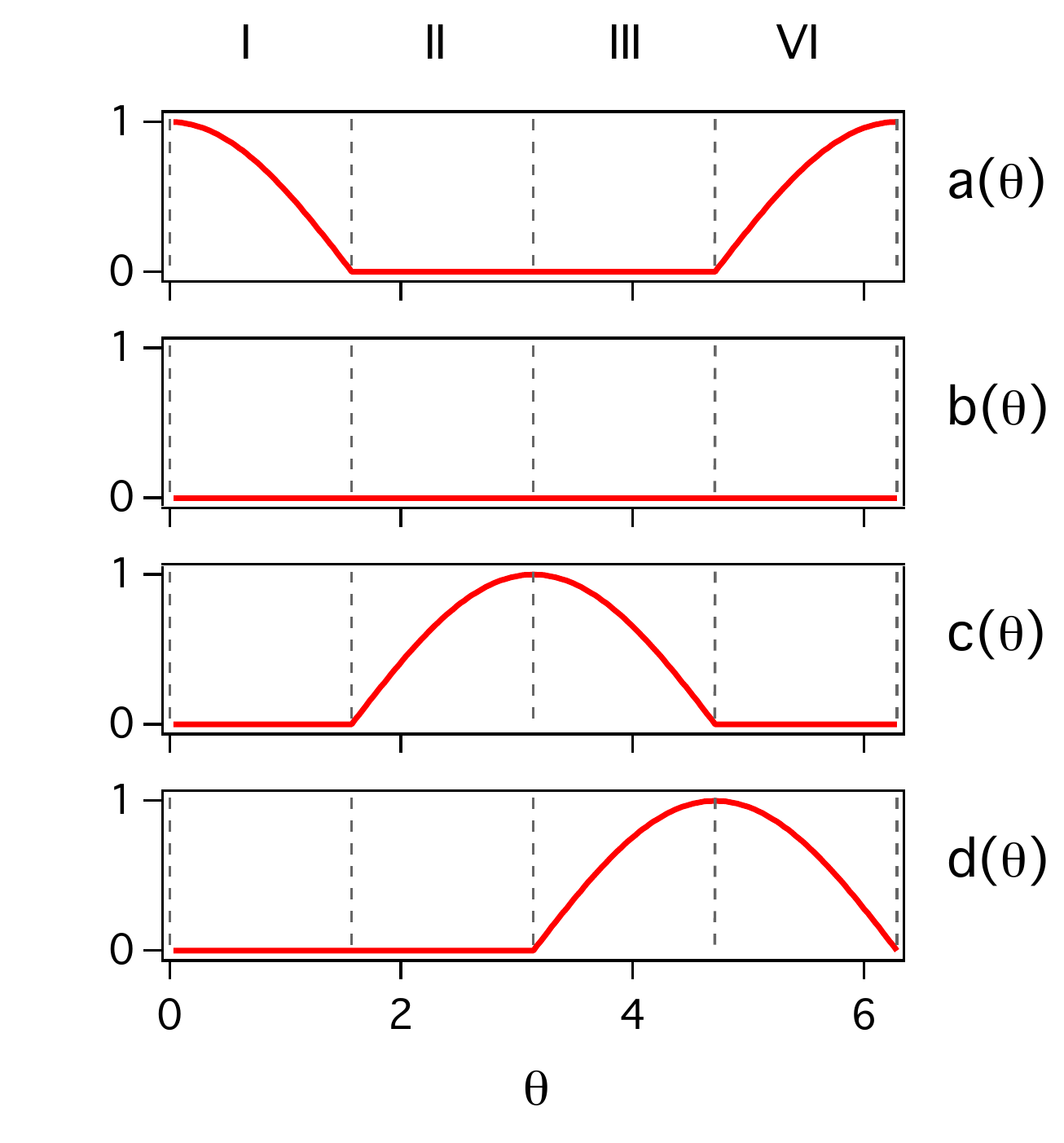}}
\caption
{\label{fig4}
Graphs of functions defined by~\eqref{shcyc} for $\theta\in[0,2\pi]$.
}
\end{figure}

\begin{figure}[h]
\center\includegraphics[width=4.5cm]{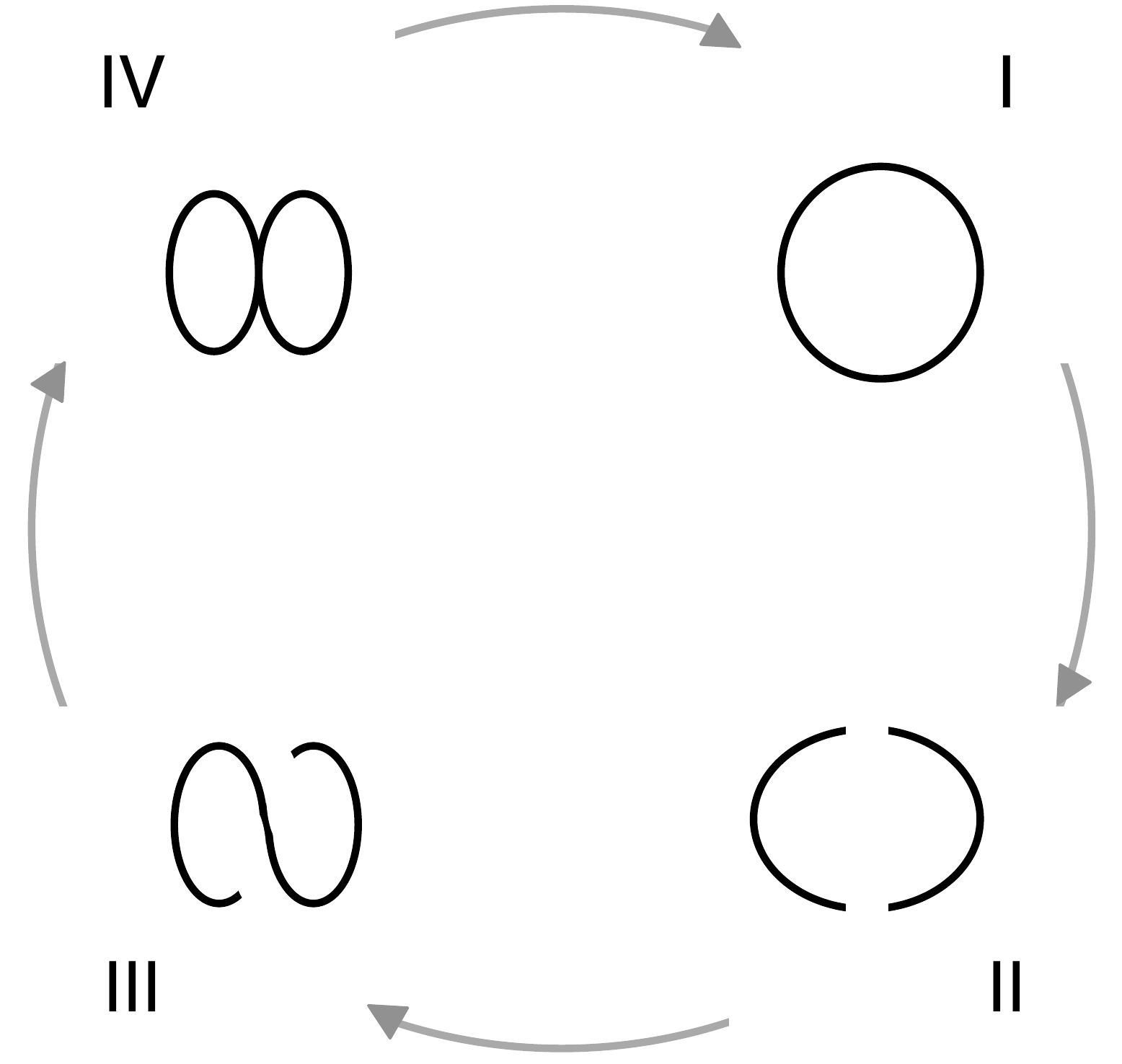}
\caption
{\label{fig5}
Cycle of topological changes corresponding to functions~\eqref{shcyc}.
}
\end{figure}

As an additional model, we consider a simplified parametric cycle represented by the boundary condition~\eqref{bc} with
\begin{eqnarray}
\label{shcyc}
&&\!\!\!\!\!\!\!\!\!\!\!\!\!\!\!\!
a(\theta) = \frac{1}{2}\left(\cos\theta  + \left| {\cos\theta } \right|\right) ,
\nonumber \\
&&\!\!\!\!\!\!\!\!\!\!\!\!\!\!\!\!
b(\theta) = b'(\theta) = 0,
\nonumber \\
&&\!\!\!\!\!\!\!\!\!\!\!\!\!\!\!\!
c(\theta) = \frac{1}{2}\left(-\cos\theta  + \left| {\cos\theta } \right|\right) ,
\nonumber \\
&&\!\!\!\!\!\!\!\!\!\!\!\!\!\!\!\!
d(\theta) = \frac{1}{2}\left(-\sin\theta  + \left| {\sin\theta } \right|\right) .
\end{eqnarray} 
The graphs of these functions are plotted in Fig.~\ref{fig4}.
This system traverses four distinct segments in the parameter space of $\theta$, as $\theta$ runs from $0$ to $2\pi$. The segments correspond to four topologies, cf. Fig.~\ref{fig5}. Namely, a ring, two disjoint lines, a line attached to a ring, and two rings connected at a point.
 
%

\section{Quantum anholonomy}

\begin{figure}[h]
\center{
\includegraphics[width=4.9cm]{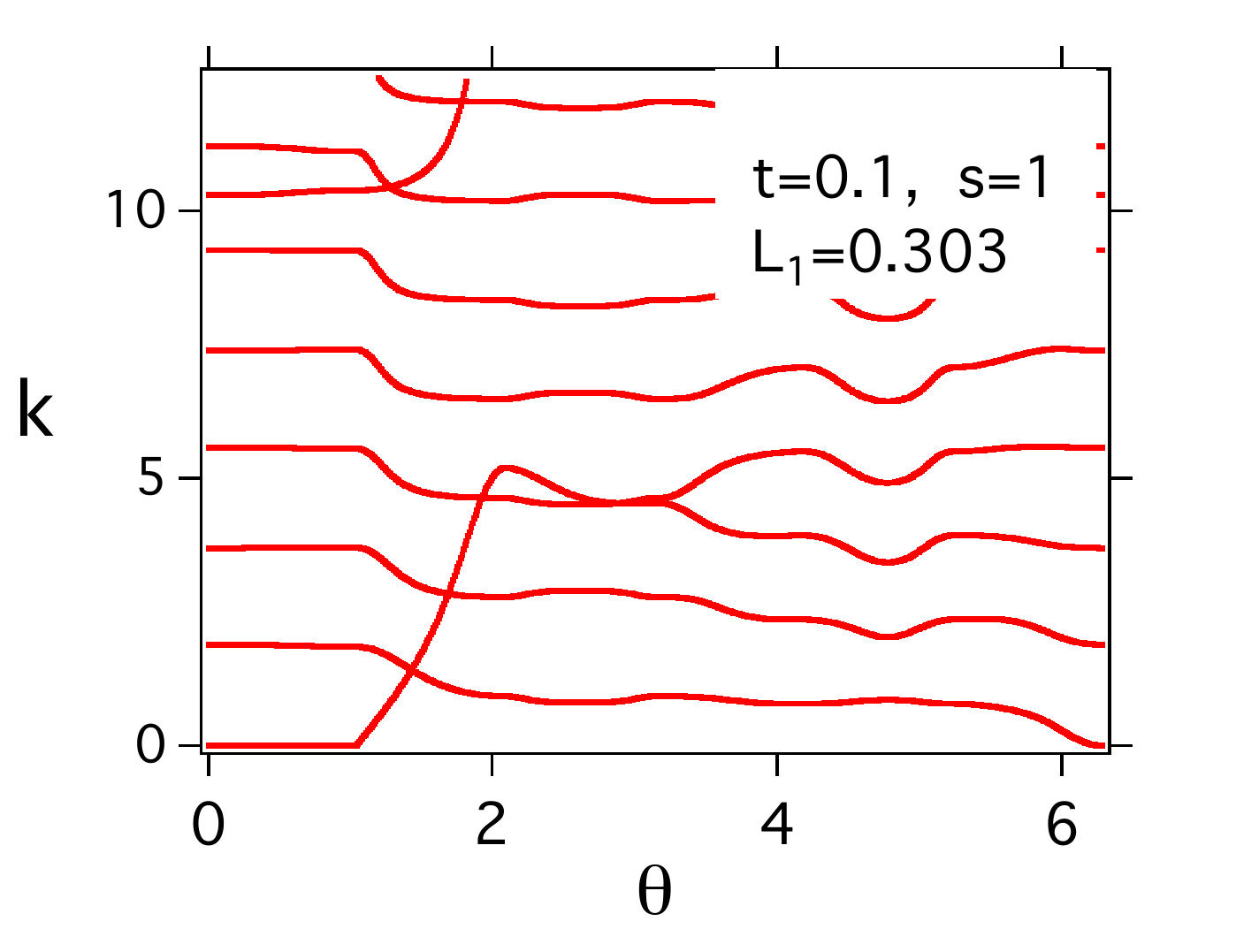} \\
\includegraphics[width=4.9cm]{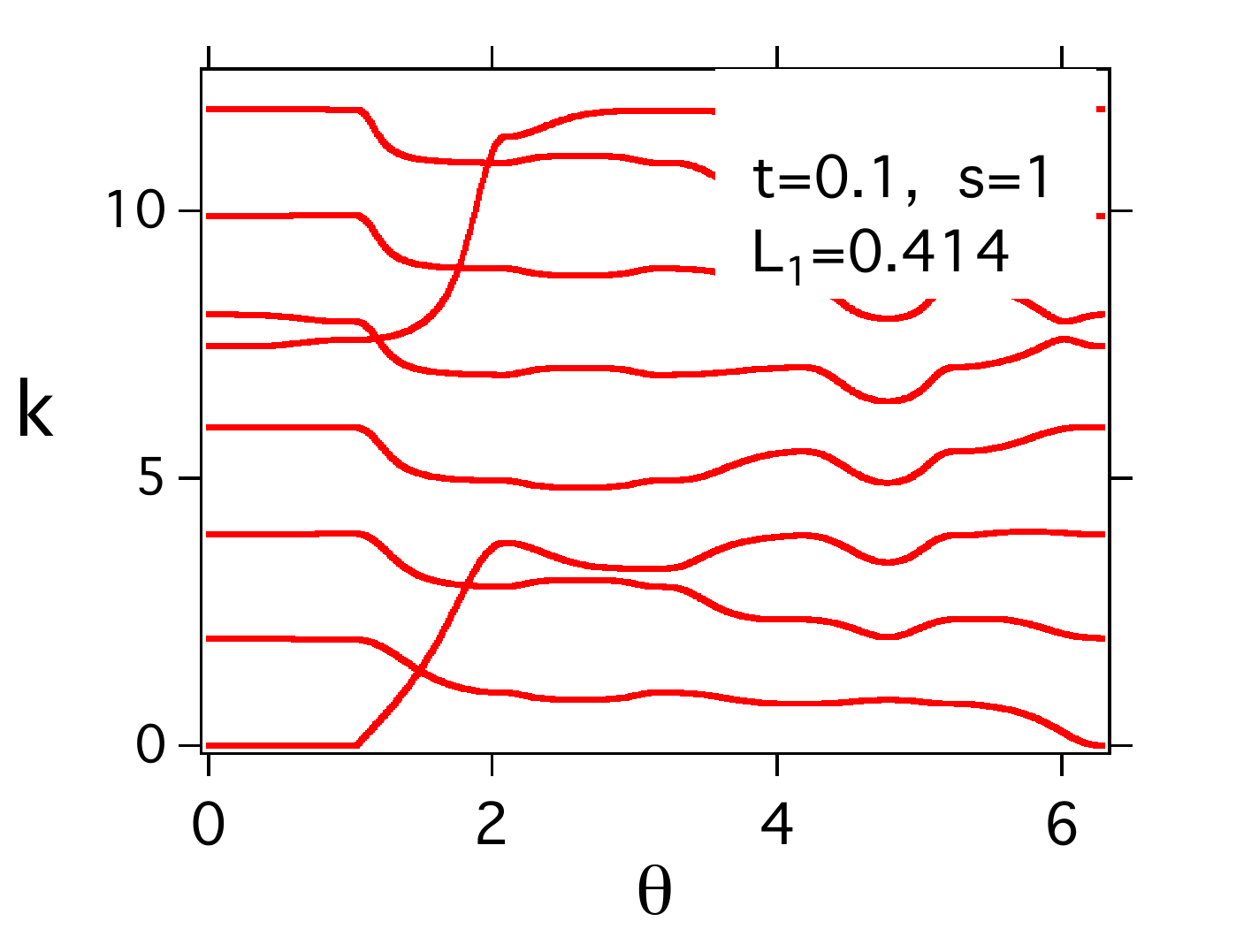} \\
\includegraphics[width=4.9cm]{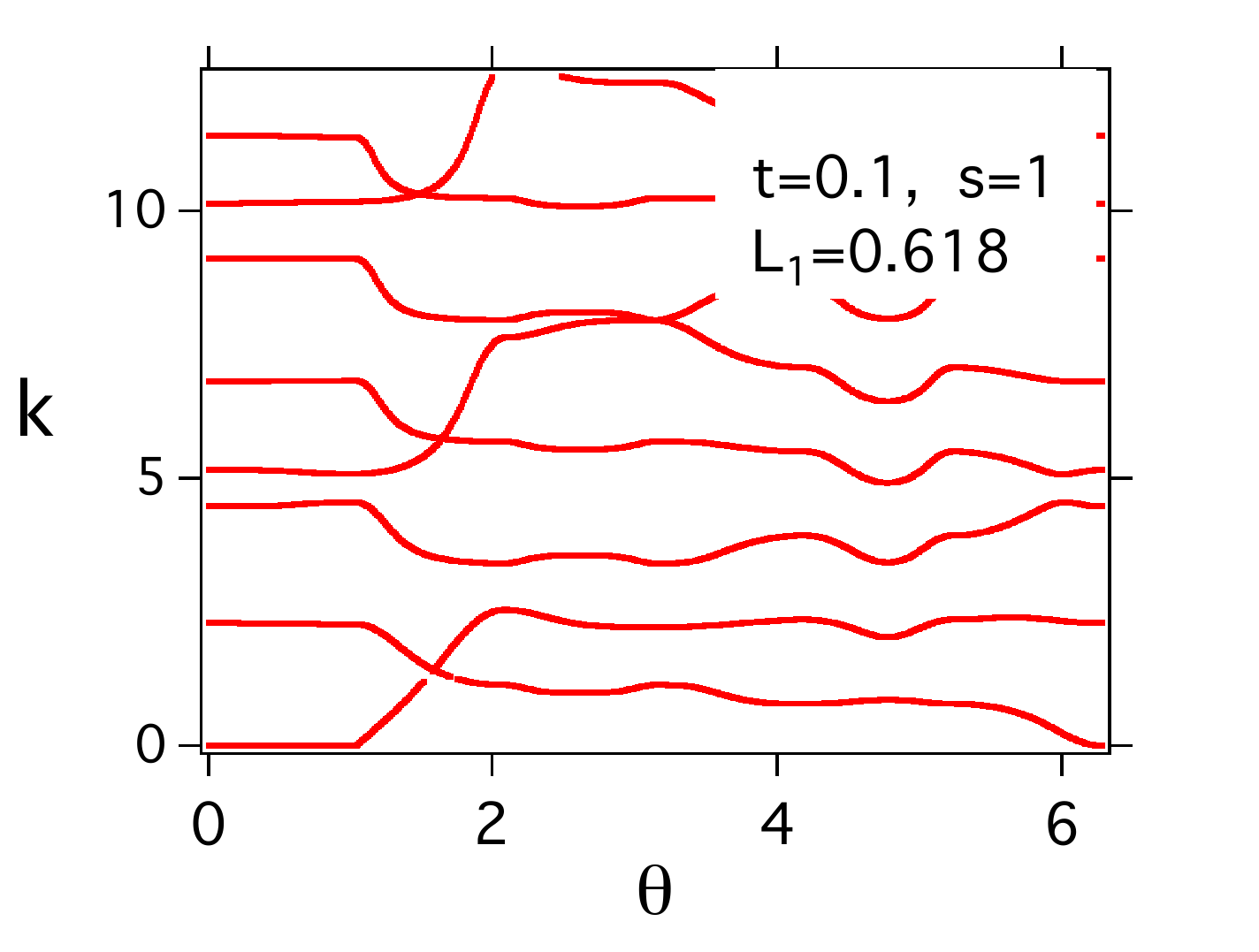}}
\caption
{\label{fig6}
Anholonomy as a result of the long cycle topology change for $t=0.1$, $s=1$. The ratio $L_1/L_2$ is chosen as the bronze mean (top), silver mean (middle) and golden mean (bottom).
}
\end{figure}
\begin{figure}[h]
\center{
\includegraphics[width=4.9cm]{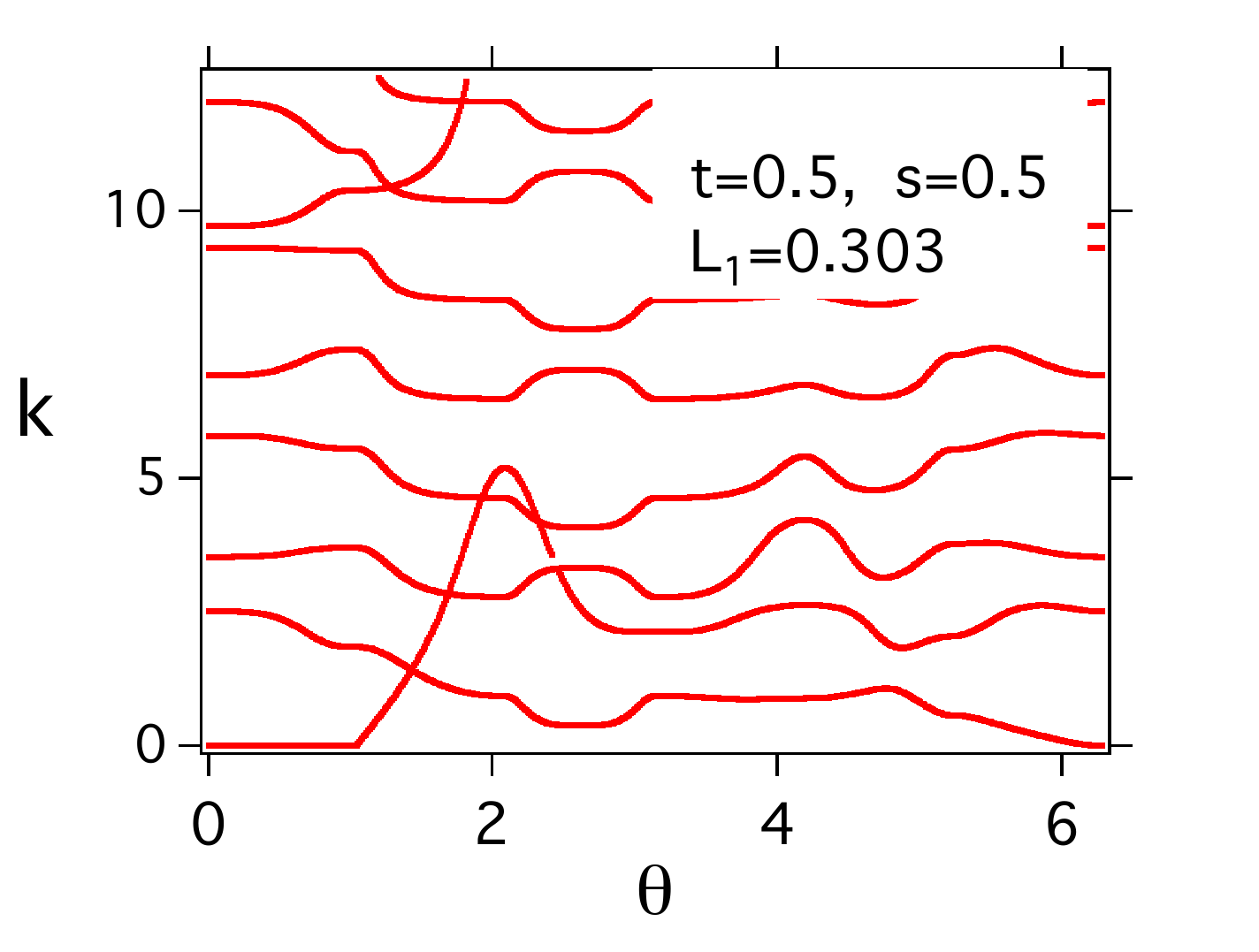} \\
\includegraphics[width=4.9cm]{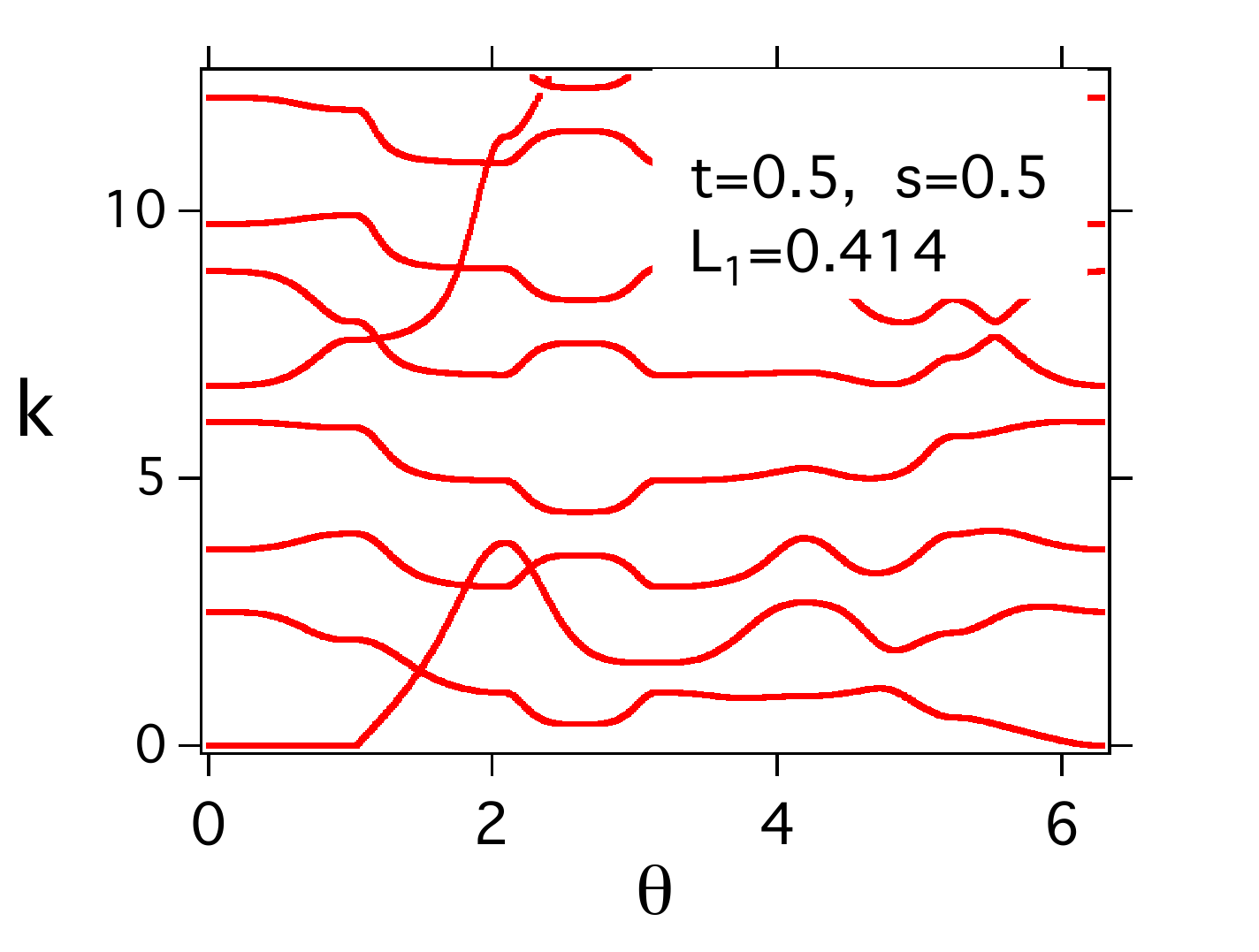} \\
\includegraphics[width=4.9cm]{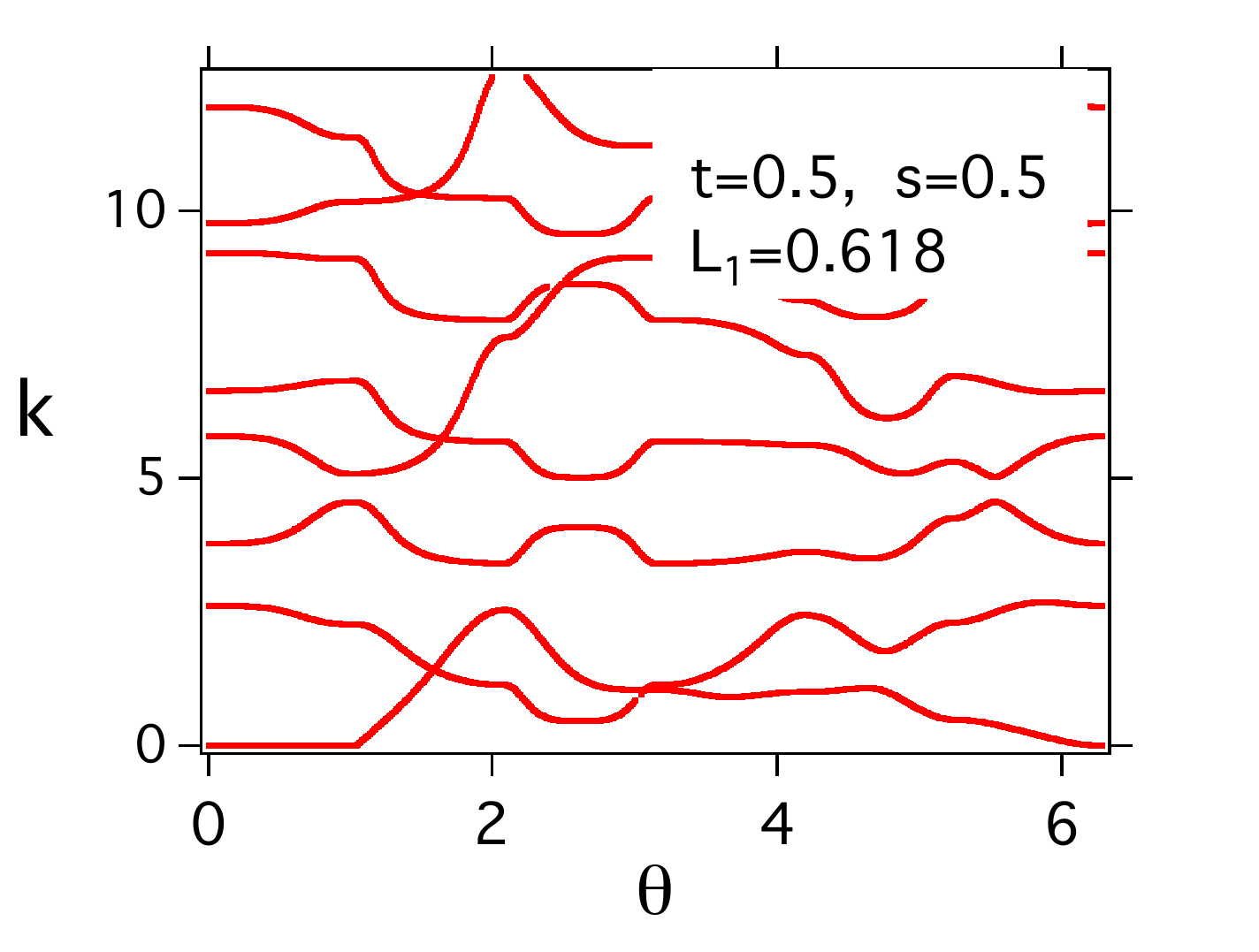}}
\caption
{\label{fig7}
Anholonomy as a result of the long cycle topology change for $t=0.5$, $s=0.5$. The ratio $L_1/L_2$ is chosen as the bronze mean (top), silver mean (middle) and golden mean (bottom).
}
\end{figure}
\begin{figure}[h]
\center{
\includegraphics[width=4.9cm]{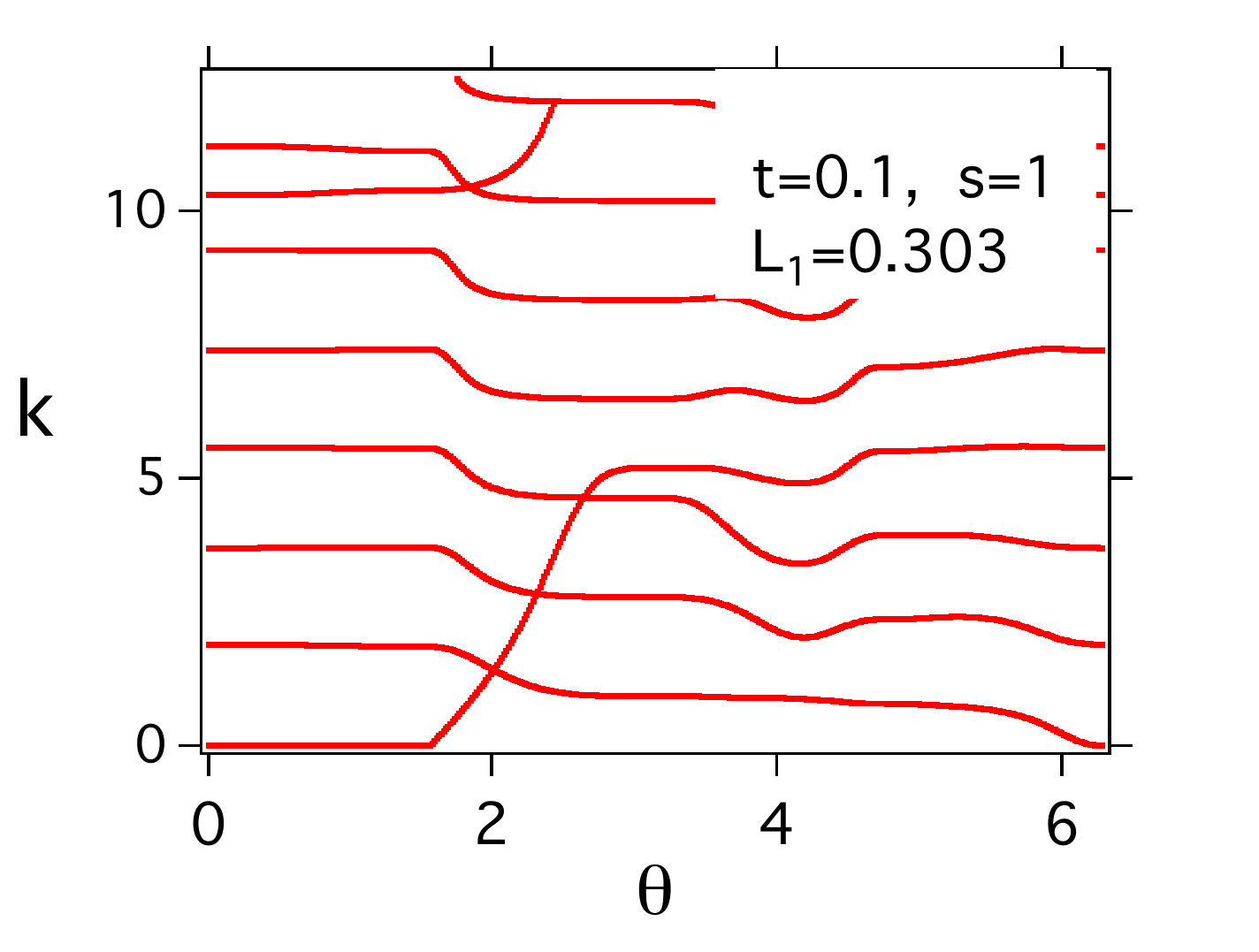} \\
\includegraphics[width=4.9cm]{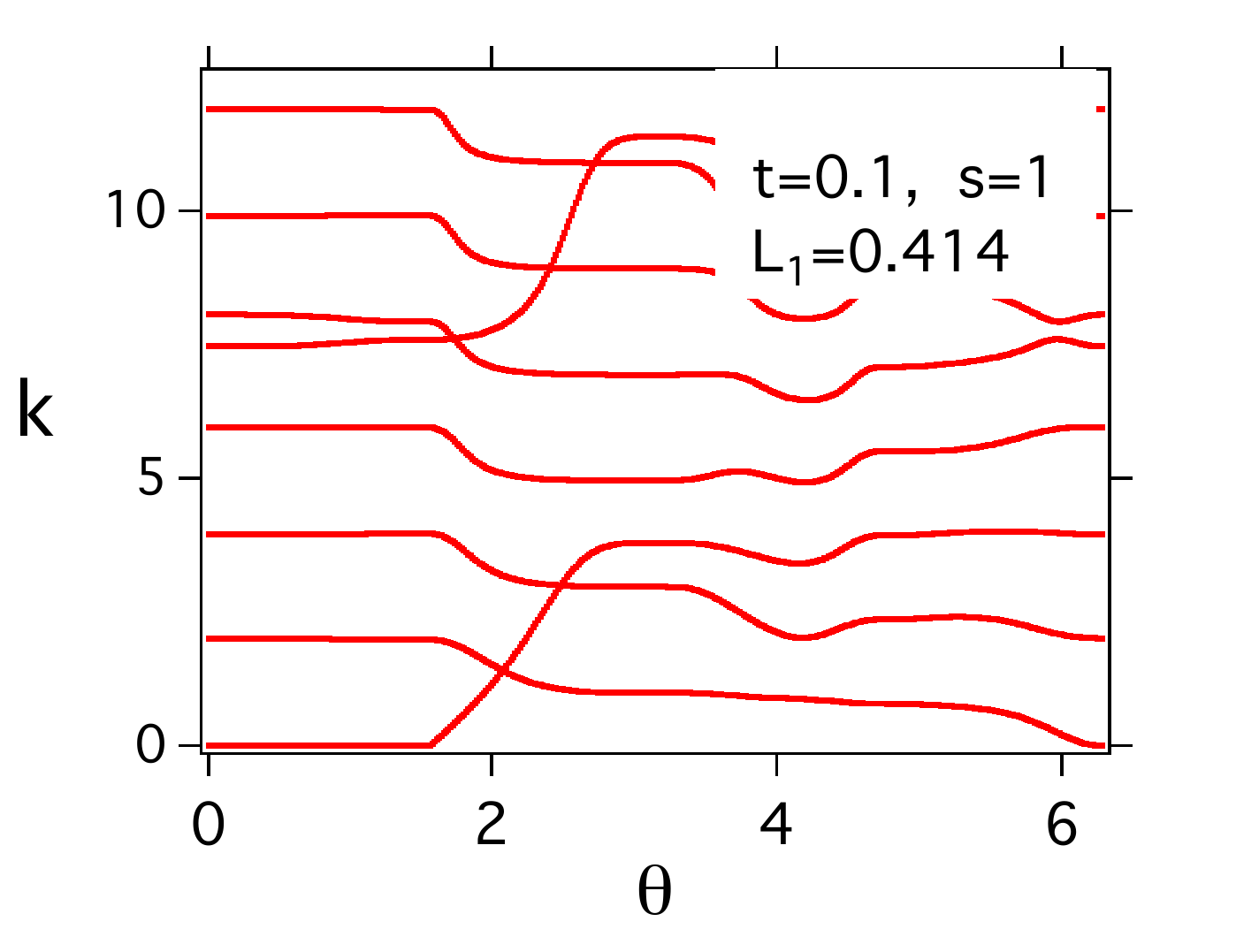} \\
\includegraphics[width=4.9cm]{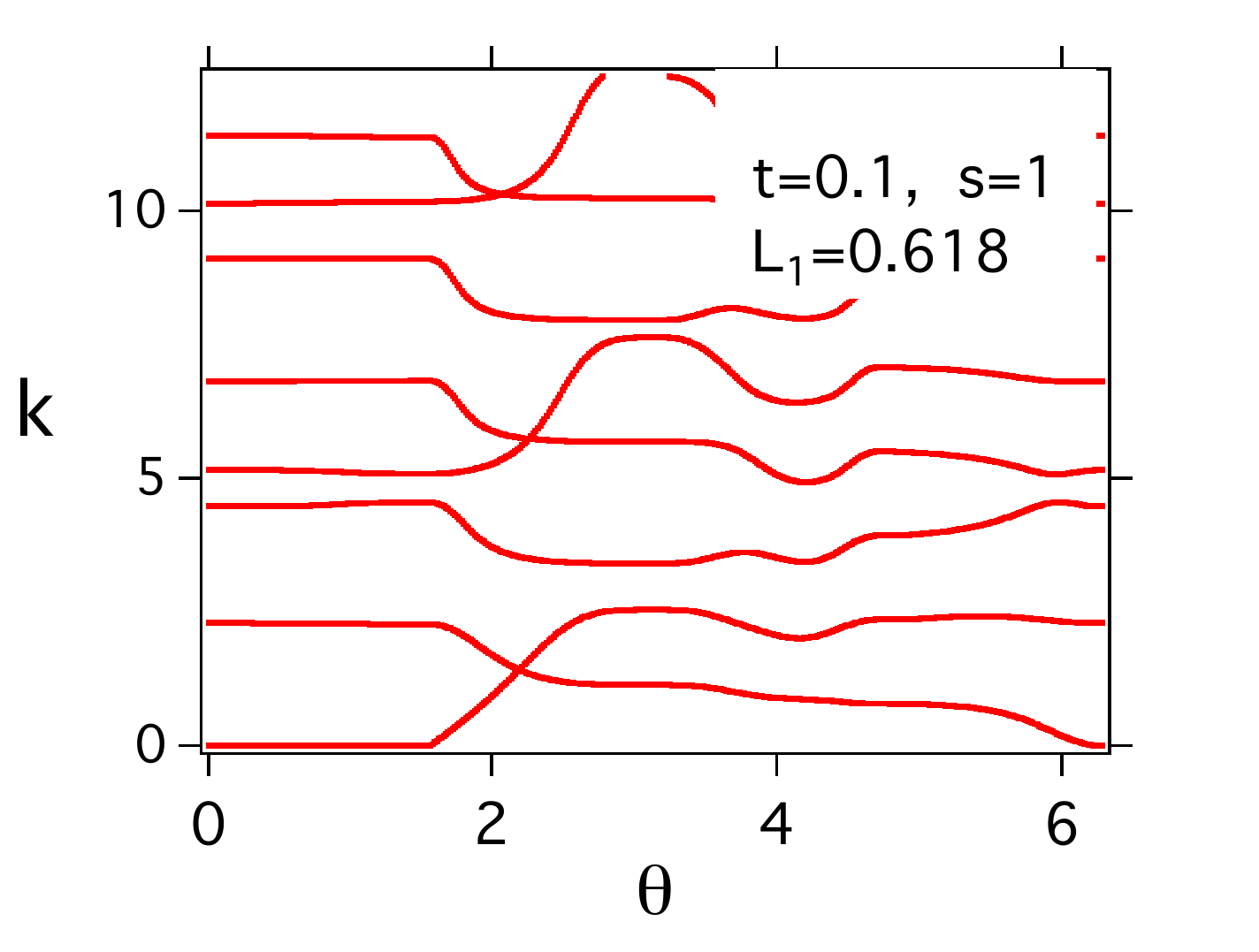}}
\caption
{\label{fig9}
Anholonomy as a result of the short cycle topology change for $t=0.5$, $s=0.5$. The ratio $L_1/L_2$ is chosen as the bronze mean (top), silver mean (middle) and golden mean (bottom).
}
\end{figure}

%
The system wave functions can be written in the form
\begin{eqnarray}
\psi_1(x_1) = \alpha_1 \sin k x_1 + \beta_1 \cos k x_1,
\nonumber \\
\psi_2(x_2) = \alpha_2 \sin k x_2 + \beta_2 \cos k x_2.
\end{eqnarray} 
In this notation, we have
\begin{eqnarray}
\Psi = U \begin{pmatrix} \alpha_1 \\ \beta_1 \\ \alpha_2 \\ \beta_2 \end{pmatrix}, 
\quad
\Psi' = k V \begin{pmatrix} \alpha_1 \\ \beta_1 \\ \alpha_2 \\ \beta_2 \end{pmatrix}
\end{eqnarray} 
with
\begin{eqnarray}
 U = \begin{pmatrix}  & 1 &  & \\ \sin k L_1 & \cos k L_1 &  &   \\  &  &  & 1 \\  & &  \sin k L_2 & \cos k L_2   \end{pmatrix} ,
 \nonumber \\
 V = \begin{pmatrix} 1 &  &  & \\ -\cos k L_1 & \sin k L_1 &  &   \\  &  & 1 &   \\  & &  -\cos k L_2 & \sin k L_2   \end{pmatrix} .
\end{eqnarray} 
The spectra of the system can be calculated from the secular equation
\begin{eqnarray}
|AU + k B V | = 0 .
\end{eqnarray} 
It turns out that certain choices of $s,t$ and ratio $L_1/L_2$ result in a quantum anholonomy. Below we present the results, separately for the long and the short cycle.


\subsection{Long Cycle}

We first consider the long cycle of parameter change represented by (\ref{locyc}).
We performed a numerical calculation of the spectra of the system for six different settings of the parameters $s,t,L_1/L_2$. We studied two combinations of $s,t$, namely
\begin{itemize}
\item $t=0.1$, $s=1$,
\item $t=0.5$, $s=0.5$,
\end{itemize}
and for each of them, we considered three different ratios $L_1/L_2$ (with $L_1+L_2=1$), namely
\begin{itemize}
\item the golden mean $L_1/L_2=(1+\sqrt{5})/2$,
\item the silver mean $L_1/L_2=1+\sqrt{2}$,
\item the bronze mean $L_1/L_2=(3+\sqrt{13})/2$.
\end{itemize}


We explain the numerical result shown in Figure \ref{fig6}. The spectra have period $2\pi$ as a function of $\theta$. This is because all the Hamiltonians are $2\pi$ periodic by construction. On the other hand, however, some wavenumber do not have the $2\pi$ periodicity. For example, let us keep track of the wavenumber (say $k_0$) of the ground state at $\theta=0$ in the uppermost figure. Around $\pi/3 < \theta < 2\pi/3$, $k_0$ increase suddenly and form three crossings with another eigenstates. Because the graphs consist of two disjoint parts, these crossings are not avoided. The eigenfunction corresponding to $k_0$ is localized within a disjoint fragment, and the other three eigenfunctions involving the crossings are localized in the other part. Hence, as a result of a $2\pi$-cycle of $\theta$, $k_0$ becomes the wavenumber of the third excited state. This is an example of anholonomy in eigenvalue and eigenspace. Some other wavenumbers, e.g., 4-th excited state at $\theta=0$ in the leftmost figure, are not involved by the anholonomy. By changing the ratio $L_1/L_2$, we see various patterns of the quantum anholonomy. Also, the anholonomies appear in various choice of the parameters, cf. Figure \ref{fig7}.



\subsection{Short Cycle}
Let us proceed to the simplified cycle represented by (\ref{shcyc}), in which four distinct topologies are connected in a single sequence.  The spectra as functions of the angle parameter $\theta \in [0, 2\pi)$ have been calculated for the choice of parameters $s=0.5$, $t=0.5$ and $L_1/L_2$ attaining the value of the golden mean, the silver mean and the bronze mean. The results are depicted in Fig.~\ref{fig9}.

\section{On the boundary condition}

\begin{figure}[h]
\center{
\includegraphics[width=2.5cm]{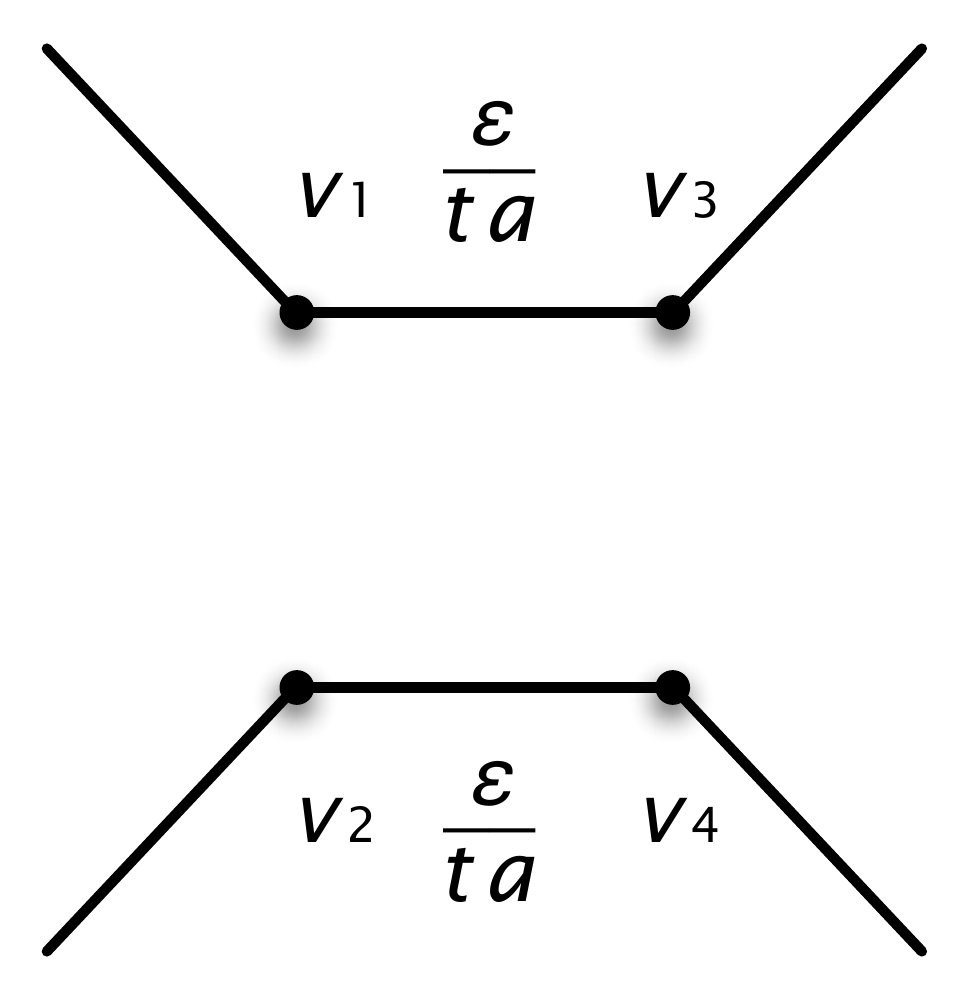} \qquad
\includegraphics[width=2.5cm]{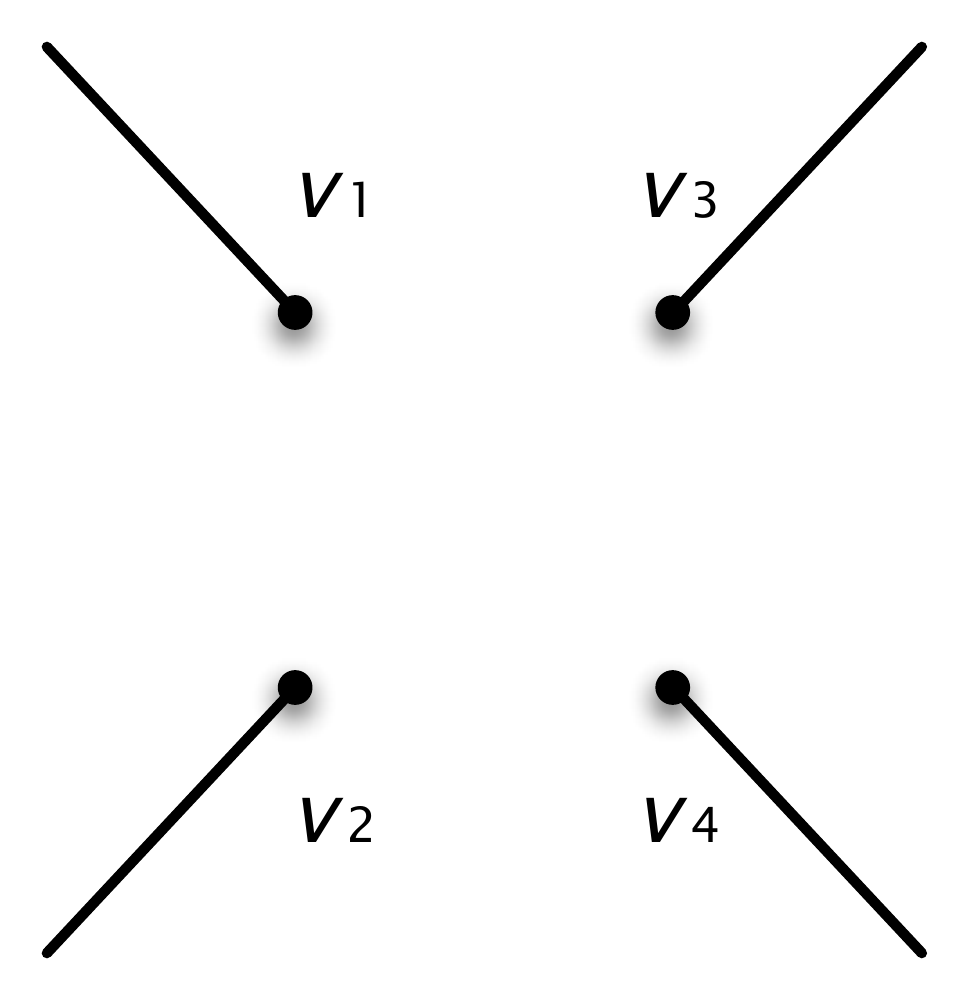}}
\caption
{\label{fig10}
Finite approximation of the physical meaning of the boundary condition~\eqref{bc} for $\theta$ belonging to the sectors I and II. The strengths of the $\delta$ potentials in the vertices 1,2,3,4 are given by the following formulas. Left: $v_1=v_2=\frac{(ta)^2-ta}{\epsilon}$, $v_3=v_4=\frac{1-ta}{\epsilon}$. Right: $v_1=0$, $v_2=\frac{c}{1-c}$, $v_3=\frac{1-c}{c}$, $v_4=\frac{1}{\epsilon}$.
}
\end{figure}

\begin{figure}[h]
\center{
\includegraphics[width=2.5cm]{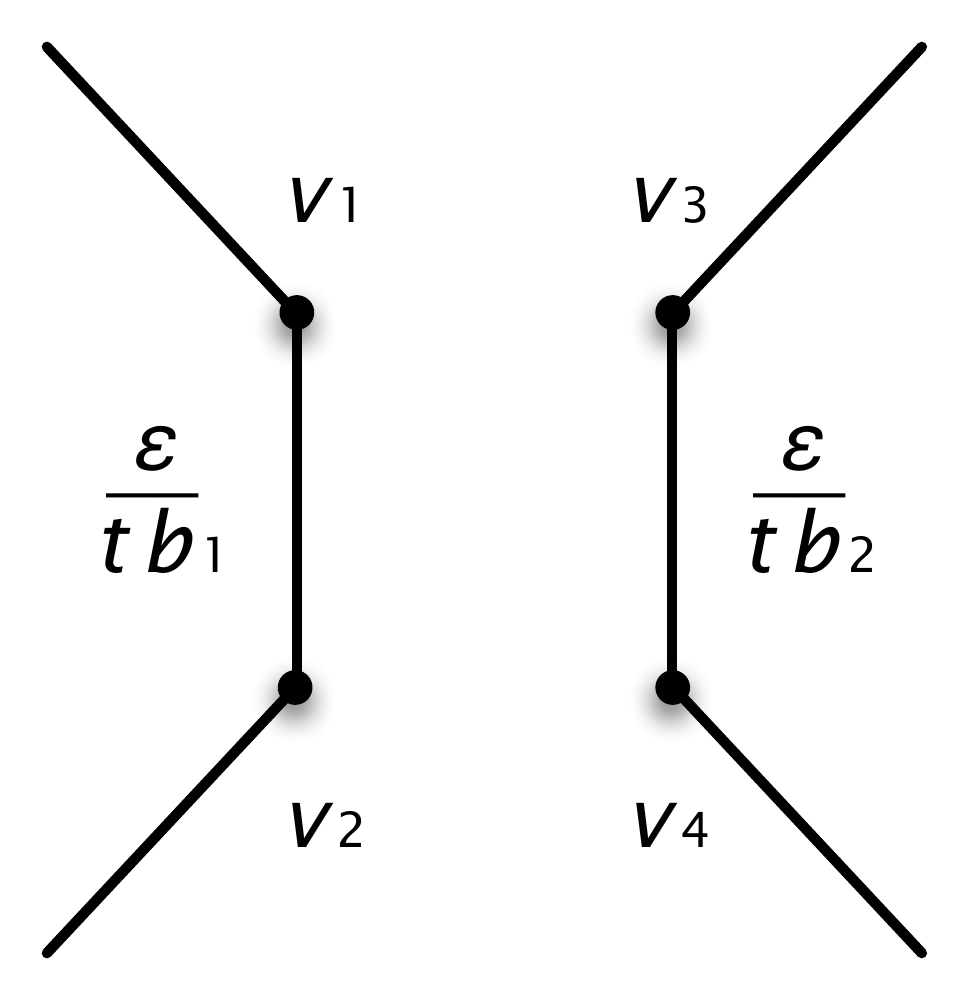} \qquad
\includegraphics[width=2.5cm]{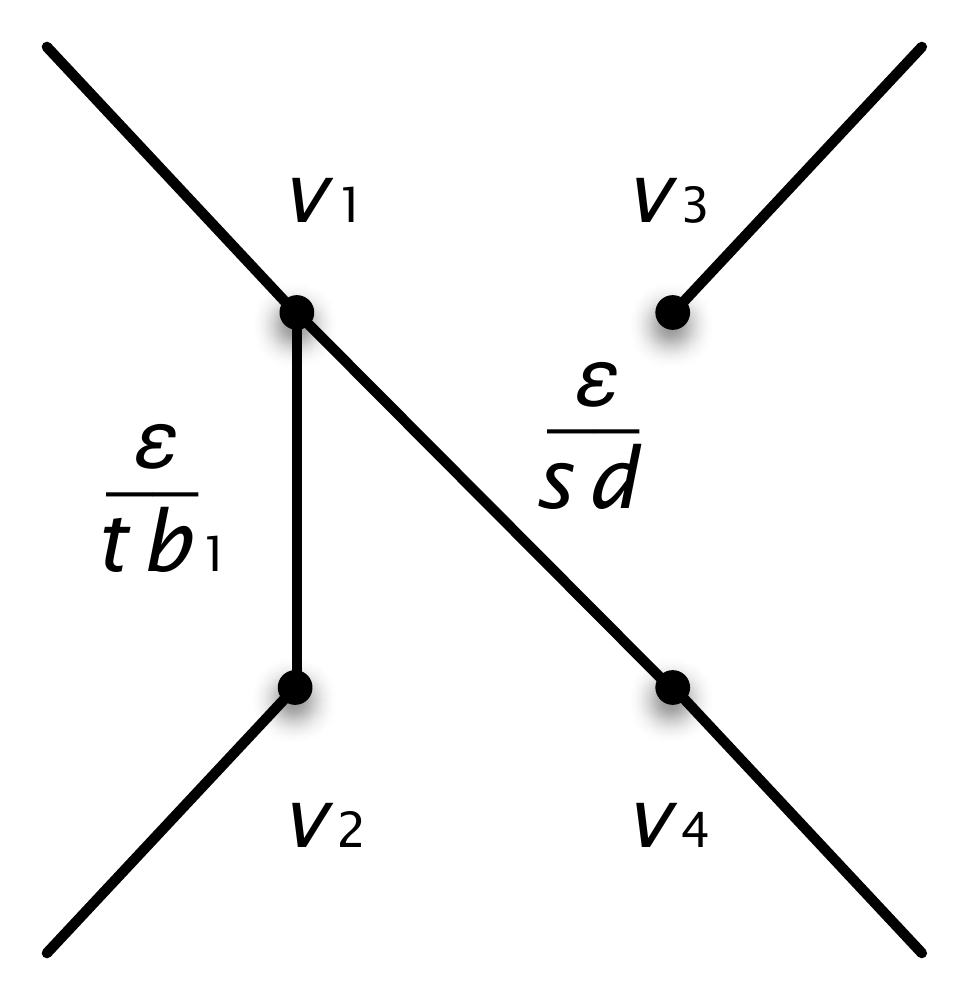}}
\caption
{\label{fig11}
Finite approximation of the physical meaning of the boundary condition~\eqref{bc} for $\theta$ belonging to the sectors III and IV. The strengths of the $\delta$ potentials in the vertices 1,2,3,4 are given by the following formulas. Left: $v_1=\frac{(tb)^2-tb}{\epsilon}$, $v_2=\frac{1-tb}{\epsilon}$, $v_3=\frac{(tb')^2-tb'}{\epsilon}$, $v_4=\frac{1-tb'}{\epsilon}$. Right: $v_1=\frac{(tb_1)^2+(sd)^2-tb_1-sd}{\epsilon}$, $v_2=\frac{1-tb_1}{\epsilon}$, $v_3=\frac{1}{\epsilon}$, $v_4=\frac{1-sd}{\epsilon}$.
}
\end{figure}

\begin{figure}[h]
\center{
\includegraphics[width=2.5cm]{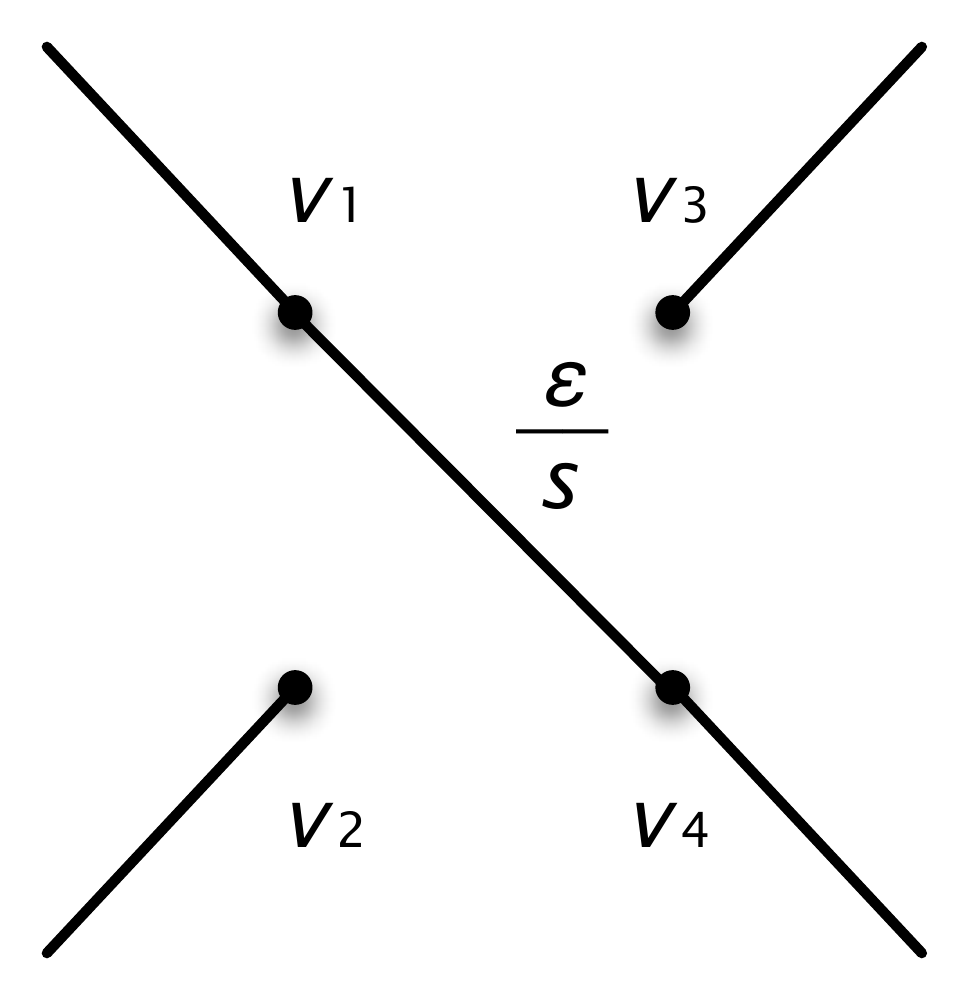} \qquad
\includegraphics[width=2.5cm]{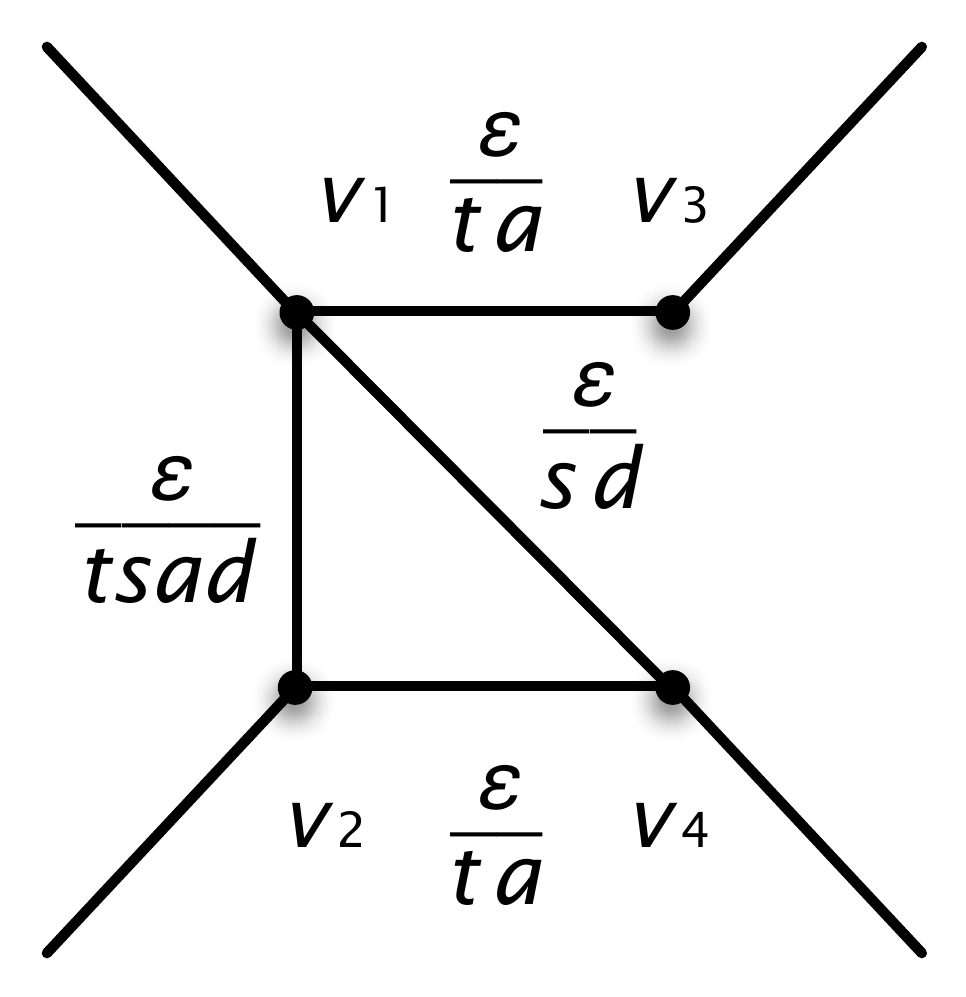}}
\caption
{\label{fig12}
Finite approximation of the physical meaning of the boundary condition~\eqref{bc} for $\theta$ belonging to the sectors V and VI. The strengths of the $\delta$ potentials in the vertices 1,2,3,4 are given by the following formulas. Left: $v_1=\frac{s^2-s}{\epsilon}$, $v_2=\frac{c}{1-c}$, $v_3=\frac{1-c}{c}$, $v_4=\frac{1-s}{\epsilon}$. Right: $v_1=\frac{(ta)^2+(sd)^2-ta-sd-3tsad}{\epsilon}$, $v_2=\frac{(ta)^2-ta-3tsad}{\epsilon}$, $v_3=\frac{1-ta}{\epsilon}$, $v_4=\frac{1-ta-sd}{\epsilon}$.
}
\end{figure}

The boundary condition~\eqref{ee1} represents a certain singular potential in the vertex that has in general a highly nontrivial nature. As it has been shown in \cite{CET10,TC13}, the physical content of a generic boundary condition corresponds to a set of infinitely strong $\delta$ potentials sited in points in an infinitely small web located in the vertex. We sketch how the webs look like for the boundary conditions of our model in Figures \ref{fig10}, \ref{fig11} and \ref{fig12}. In all of the finite approximations, the parameter $\epsilon$ has to be chosen small with respect to the wavelength of the particle. For $\epsilon\to0$, the approximation effectively produces the boundary conditions~\eqref{bc}. The result has been obtained using the formulas from \cite{CET10,TC13}. It illustrates in a simple way the relation between the global topology of the system and the topology of the internal infinitely small web.

\section{Discussion}

The creation of the anholonomy in our model can be explained in the following way.
In regions I, IV, V and VI, when the graph consists of a single component, the eigenvalues as functions of the system parameter $\theta$ can change their values, but they never cross each other, because generically there will be avoided crossings.  However, in regions II and III, the graph is separated into two components (two lines or two rings), i.e., the system is in fact made up of two independent entities, both having its own energy levels that can independently vary with change of the system parameter $\theta$. Therefore, if the parametrization of boundary conditions is conveniently chosen, the eigenvalues as functions of $\theta$ can cross each other, which leads to the anholonomy effect.

\begin{acknowledgements}
We thank Professor P. Exner and Professor I. Tsutsui for stimulating discussions. This research was supported  by the Japan Ministry of Education, Culture, Sports, Science and Technology under the Grant numbers 24540412 and 22540396.
\end{acknowledgements}

\end{document}